\documentclass[aps,preprint,showpacs,floats,epsf,epsfig,nofootinbib,12pt]{revtex4}
\usepackage{graphicx}
\usepackage{epsfig}
\usepackage[dvips]{color}

\def\be{\begin{eqnarray}}
\def\en{\end{eqnarray}}

\begin{document}

\title{Implications of equalities among the elements of CKM and PMNS matrices}

\vspace{1cm}

\author{ Hong-Wei Ke$^1$\footnote{khw020056@hotmail.com}, Song-Xue Zhao$^1$ and
        Xue-Qian Li$^2$\footnote{lixq@nankai.edu.cn}  }

\affiliation{  $^{1}$ School of Science, Tianjin University, Tianjin 300072, China \\
  $^{2}$ School of Physics, Nankai University, Tianjin 300071, China }

\vspace{12cm}

\begin{abstract}
Investigating the CKM matrix in different parametrization schemes,
it is noticed that those schemes can be divided into a few groups where
the sine values of the CP phase for each group are  approximately
equal. Using those relations, several approximate equalities among
the elements of CKM matrix are established. Assuming them to be
exact, there are infinite numbers of solutions  and by choosing
special values for the free parameters in those solutions, several textures presented
in literature are obtained. The case can also be generalized to
the PMNS matrix for the lepton sector. In parallel, several mixing textures are also derived
by using presumed  symmetries, amazingly, some of their
forms are the same as what we obtained, but not all. It hints
existence of a hidden symmetry which is broken in the practical
world. The nature makes its own selection on the underlying
symmetry and the way to break it, while we just guess what it is.

\pacs{12.15.Ff, 14.60.Pq, 12.15.Hh}

\end{abstract}

\maketitle

\section{Introduction}
Due to the mismatch between the eigenstates of weak interaction
and that of mass, the $3\times 3$ unitary
Cabibbo-Kobayashi-Maskawa (CKM) matrix
\cite{Cabibbo:1963yz,Kobayashi:1973fv} is introduced to mix the
three generation quarks
\cite{Fritzsch:1979zq,Li:1979zj,Fritzsch:1997fw}, which is
determined by three independent mixing angles and one CP-phase. The CKM
matrix can be parametrized in different schemes and there are nine
schemes proposed in literatures.
Generally, the values of the three angles and CP phase can be
different for various paramtrization schemes.  By
closely investigating the matrix, it is noticed that there exist
some relations\cite{Ke:2014gda} among the CP phases in these
schemes. For convenience let us label the nine schemes with
subscripts  $a$ through $i$. Namely,  we may divide the nine
parametrization schemes into a few groups and determine
corresponding equalities among those as $\sin\delta_n$ i.e. $
\sin\delta_a\approx \sin\delta_d\approx \sin\delta_e,\,
\sin\delta_b\approx \sin\delta_c,\, \sin\delta_f\approx
\sin\delta_h\approx \sin\delta_i$. Then considering constraint of
the Jarlskog invariant\cite{Jarlskog:1985ht}, the above relations
lead to several approximate equalities among the CKM matrix elements
$|U_{jk}|$ which are measured in experiments. These equalities are indeed
approximate, but independent of any concrete
parametrization scheme.

These equalities tempt us to guess that there should exist
underlying symmetries to determine them\cite{Ke:2014gda}. Our discussion on the implications of
these equalities is based on
observation and phenomenological. In parallel,
an alternative route  was also suggested  that these
equalities can be deduced by rephasing the invariants of quark
mixing matrix\cite{Jenkins:2007ip} as long as the mixing angles among quarks being small.
In order to clarify the physical picture we would further study these equalities.

In analog to the quark sector, the Pontecorvo-Maki-Nakawaga-Sakata
(PMNS) matrix \cite{Maki:1962mu,Pontecorvo:1967fh} relates the
lepton flavor eigenstates with the mass eigenstates. Thus it is
natural to extend the relations for the CKM matrix to the PMNS
case. It is not a surprise, we find that all the equalities also hold for the lepton
sector, even though the accuracy is not as high as for the quark
sector. The allegation based on only rephasing \cite{Jenkins:2007ip} is
incomplete because it cannot explain why these
equalities also hold for neutrino mixing where at least two
mixing angles are large.

Since these equalities are respected by both CKM and PMNS, it is tempted to
conjecture that there might be an underlying symmetry to
result in the symmetric forms for both CKM and PMNS matrices which are
broken in the practical world. Based on the group
theory Lam showed a possibility that the mixing matrices originate
from a higher symmetry\cite{Lam:2011ip} which then breaks differently for
quark and lepton sectors.
The existence of the
quark-lepton complementarity and
self-complementarity\cite{Minakata:2004xt,Raidal:2004iw,Altarelli:2009kr,
Zheng:2011uz,Zhang:2012pv,Zhang:2012zh,Haba:2012qx,Ke:2014hxa}
also hints a higher symmetry. All the
progress in this area inspires a trend of searching for whether
such a high symmetry indeed exists and moreover investigation of
its phenomenological implication is also needed.

Following this idea, we assume that the equalities are exact to compose equations,
solving the equations, these solutions might offer hints towards the unknown symmetry. To confirm or testify
the scenario, we further investigate the
implication of these resultant matrices. It is found that
these solutions coincide with the symmetrical CKM and PMNS textures.
Moreover, some authors recently reached some symmetric textures based on presumed symmetries,
and it is found that some of their resultant forms are the same as ours, but not all.
We will further discuss the implications in the last section.

The paper is organized as follows. After the introduction we
review those equalities in section II. In section III, we present
the solutions which satisfy those equalities (in fact, a few
groups of solutions, and each of them contains a free parameter)
and their implications. In section IV we make a summary and
discussion.

\section{Relations among elements of the CKM matrices}

Mixing among different flavors of quarks via the CKM matrix has
been firmly recognized and the $3\times 3$ mixing matrix is
written as
    \begin{equation}\label{M1}
      V=\left(\begin{array}{ccc}
        V_{11} &V_{12} &V_{13} \\
         V_{21} &   V_{22} &  V_{23}\\
          V_{31} & V_{32} & V_{33}
      \end{array}\right).
  \end{equation}

Generally, for a $3\times 3$ unitary matrix there are four
independent parameters, namely three mixing angles and one
CP-phase. There can be various schemes to parameterize the matrix
and only nine  schemes are independent which are clearly listed in
Ref.\cite{Zhang:2012pv}. For readers' convenience, we collect them in Tab.
\ref{tab1}.

\begin{table*}
      \caption{Nine different parametrization schemes for CKM matrix}\label{tab1}
      \begin{tabular}{cccc}\hline\hline
      Scheme &  Jarlskog invariant& CP phase\\
      \hline
      ${\rm P}_{a}$ &
      $J_a=s_{a1}s_{a2}s_{a3}c_{a1}c_{a2}c^2_{a3}\sin\delta_a$ &$\delta_a=\left(69.10^{+2.02}_{-3.85}\right)^\circ$
      \\
      ${{\rm P}_{b}}$ &
      $J_b=s_{b1}s^2_{b2}s_{b3}c_{b1}c_{b2}c_{b3}\sin\delta_b$&$\delta_b=\left(89.69^{+2.29}_{-3.95}\right)^\circ$
        \\
      ${{\rm P}_{c}}$ &
      $J_c=s^2_{c1}s_{c2}s_{c3}c_{c1}c_{c2}c_{c3}\sin\delta_c$& $\delta_c=\left(89.29^{+3.99}_{-2.33}\right)^\circ$
      \\
      ${{\rm P}_{d}}$ &
      $J_d=s_{d1}s_{d2}s_{d3}c^2_{d1}c_{d2}c_{d3}\sin\delta_d$&$\delta_d=\left(111.95^{+3.82}_{-2.02}\right)^\circ$
             \\
      ${{\rm P}_{e}}$ &
      $J_e=s_{e1}s_{e2}s_{e3}c_{e1}c^2_{e2}c_{e3}\sin\delta_e$&$\delta_e=\left(110.94^{+3.85}_{-2.02}\right)^\circ$
      \\
      ${{\rm P}_{f}}$ &
      $J_f=s_{f1}s_{f2}s_{f3}c_{f1}c_{f2}c^2_{f3}\sin\delta_f$ &$\delta_f=\left(22.72^{+1.25}_{-1.18}\right)^\circ$
      \\
      ${{\rm P}_{g}}$ &
      $J_g=s^2_{g1}s_{g2}s_{g3}c_{g1}c_{g2}c_{g3}\sin\delta_g$&$\delta_g=\left(1.08^{+0.06}_{-0.06}\right)^\circ$
      \\
      ${{\rm P}_{h}}$ &
      $J_h=s_{h1}s_{h2}s_{h3}c_{h1}c^2_{h2}c_{h3}\sin\delta_h$&$\delta_{h}=\left(157.31^{+1.18}_{-1.25}\right)^\circ$
      \\
      ${{\rm P}_i}$ &
      $J_i=s_{i1}s_{i2}s_{i3}c^2_{i1}c_{i2}c_{i3}\sin\delta_i$&$\delta_i=\left(158.32^{+1.13}_{-1.20}\right)^\circ$
       \\\hline\hline
      \end{tabular}
\end{table*}

To be more clearly, we present the the explicit expressions of two typical parametrization schemes P$_a$ and P$_e$   as
 \begin{equation}\label{M1}
      V_{{P}_a}=\left(\begin{array}{ccc}
        c_{a1}c_{a3} & s_{a1}c_{a3} & s_{a3}\\
       -c_{a1}s_{a2}s_{a3} -s_{a1}c_{a2}e^{-i \delta_a} & -s_{a1}s_{a2}s_{a3}+c_{a1}c_{a2}e^{-i\delta_a} & s_{a2}c_{a3}\\
      -c_{a1}s_{a2}s_{a3} + s_{a1}s_{a2}e^{-i\delta_a} & -s_{a1}s_{a2}s_{a3}-c_{a1}s_{a2}e^{-i\delta_a} & c_{a2}c_{a3}
      \end{array}\right),
\end{equation}
and
\begin{equation}\label{M2}
      V_{{P}_e}=\left(\begin{array}{ccc}
 -s_{e1}s_{e2}s_{e3} +c_{e1}c_{e3}e^{-i \delta_e} & -c_{e1}s_{e2}s_{e3}-s_{e1}c_{e2}e^{-i\delta_e} & c_{e2}s_{e3}\\
        s_{e1}c_{e2} & c_{e1}c_{e2} & s_{e2}\\
            -s_{e1}s_{e2}c_{e3} - c_{e1}s_{e3}e^{-i\delta_e} & -c_{e1}s_{e2}c_{e3}+s_{e1}s_{e3}e^{i\delta_e} & c_{e2}c_{e3}
\end{array}\right).
\end{equation}
Here $s_{aj}$ and $c_{aj}$ ($s_{ej}$ and $c_{ej}$) denote
$\sin\theta_{aj}$ and $\cos\theta_{aj}$ ($\sin\theta_{ej}$ and
$\cos\theta_{ej}$) with $j=1,2,3$. $\theta_{nj}$ and $\delta_n$
are the mixing angles and CP-phase respectively. The corresponding expressions
in other schemes P$_n$ can be found in Ref.\cite{Zhang:2012pv}.

From the data measured in various experiments, one can deduce values of
the angles  $\theta_{nj}$ and CP phase $\delta_n$ which are not the same for different parametrizations.

Close observation on the values of $\delta_n$ in different
schemes exhibits several approximate equalities
\begin{eqnarray} \label{rl1}
{\rm sin}\delta_a\approx {\rm sin}\delta_d\approx {\rm
sin}\delta_e,\, {\rm sin}\delta_b\approx {\rm sin}\delta_c,\,{\rm
sin}\delta_f\approx  {\rm sin}\delta_h\approx {\rm sin}\delta_i.
\end{eqnarray}
Namely, the nine phase factors in the nine schemes are divided into a few groups and
their sine values in each group are approximately equal.
It is well known that the Jarlskog invariant is independent of
schemes, so using the above relations in Eq.(\ref{rl1})
and substituting $s_{nj}$ and $c_{nj}$ with the ratios of modules
of corresponding elements, one can deduce several interesting
relations among the elements of CKM, which are
experimentally measured values and obviously free of
parametrization schemes:

\begin{eqnarray} \label{rl4}
 \frac{{|V_{21}|} {|V_{22}|}
}{
   1-{|V_{23}|}^2}-\frac{{|V_{11}|} {|V_{12}|}
   }{
    {|V_{23}|}^2+{|V_{33}|}^2}\approx 0\nonumber\\
   \frac{{|V_{11}|}  {|V_{12}|} {|V_{21}|}}{
   1-{|V_{11}|}^2}
   -\frac{{|V_{23}|} {|V_{32}|} {|V_{33}|} }{
   1-{|V_{33}|}^2}\approx 0\nonumber\\
 \frac{{|V_{21}|} {|V_{23}|}  {|V_{33}|}}{
   1-{|V_{23}|}^2}-\frac{{|V_{11}|} {|V_{12}|}
   {|V_{32}|}}{
   {|V_{22}|}^2+{|V_{32}|}^2}\approx0\nonumber\\
 \frac{{|V_{12}|} {|V_{22}|}}{
   1-{|V_{32}|}^2}
   \frac{{|V_{11}|} {|V_{21}|} }{{|V_{11}|}^2+{|V_{21}|}^2
  }\approx0\nonumber\\
 \frac{{|V_{12}|} {|V_{32}|}  {|V_{33}|}}{{|V_{12}|}^2+{|V_{22}|}^2
   }-
   \frac{{|V_{11}|} {|V_{21}|}  {|V_{23}|}}{
   {1-|V_{21}|}^2}\approx0.
\end{eqnarray}

\section{Implication of the relations}
\subsection{On these relations}
Even though, our allegation starts from a phenomenological observation, it is
natural to attribute these equalities to an underlying symmetry.
In parallel, it was argued that they can automatically emerge
from a different ansatz which we briefly outline in the appendix.

These relations are proved to be exact equalities  under the
limit $\theta_{a2}\rightarrow0$ and  $\theta_{a3}\rightarrow0$, so they are the
consequence of small $\theta_{a2}$ and  $\theta_{a3}$ and the practical approximation indeed comes from being non-zero.
For an illustration, anyone can check those relations for P$_a$ parametrization and we present the details in Appendix A.
There, since $\theta_{a1}=(13.023^{+0.038}_{-0.038})^\circ,
\theta_{a2}=(2.360^{+0.065}_{-0.038})^\circ,
\theta_{a3}=(0.201^{+0.010}_{-0.008})^\circ $\cite{Zhang:2012pv},
the picture seems work almost perfect.

Another way to obtain these equalities can be started from
rephasing the invariants of quark mixing matrix. In
Ref.\cite{Jenkins:2007ip} the authors pointed out that
$V_{i\alpha} V^*_{j\alpha} V_{j\beta} V^*_{i\beta}$ are invariants
whose imaginary is the traditional Jarlskog invariant. Since
$V_{i\alpha} V^*_{j\alpha} V_{j\beta} V^*_{i\beta}$ are invariants
in different parametrizations one can use them to deduce relations
among the physical matrix elements. For example  by comparing the
real parts and imaginary parts of the invariant $V_{12}  V^*_{22}
V_{23}V^*_{13}$ in P$_a$ and P$_e$ parametrizations ${\rm
sin}\delta_a\approx {\rm sin}\delta_e$ can be deduced  with the
postulates of small $\theta_{a2}$ and $\theta_{a3}$.  Some details
are presented in Appendix B.

The two ways are similar as the same condition that
$\theta_{a2}$ and $\theta_{a3}$ being small is taken. If one just discusses the quark
case the two ways seem to to be parallel. However if one tries to extend these
relations to the PMNS case, he needs to reconsider them more
carefully because then the conditions of small mixing angles no longer exist.

In fact, assuming those relations to be exact, solving the equations we obtain
several independent solutions and each of them contains a free parameter to be fixed.

In the next section we will show
that for the quark sector, the two ways correspond just to special choices of the parameters in the
solutions, but for the lepton sector they are different.

\subsection{Solutions of these relations}
Now we replace the $``\approx"$ with equal sign $``="$ in
Eq.(\ref{rl4}) to compose equations and obtain their solutions. Since
these solutions are expected to correspond to the symmetrical
textures for CKM and/or PMNS matrices, the normalization of the
unitary matrix
\begin{eqnarray} \label{rl3}
{|V_{11}|}^2+{|V_{12}|}^2+{|V_{13}|}^2=1,
{|V_{11}|}^2+{|V_{21}|}^2+{|V_{31}|}^2=1, ...
\end{eqnarray}
should be retained.

It is noted that even though we establish the equalities from equating the CP phases of
different parametrizations, in the later procedures only the ratios of modules of the matrix elements are
employed to build up equations, thus  one cannot gain any information about the phases
of the matrix elements from the normalization relations and
Eq.(\ref{rl4}). If one hopes to know the phases of the elements some new
constraints must be further enforced, such as orthogonality
between any two different rows or columns of the matrix. Now, the newly built equations are free of
concrete parametrizations.


Satisfying all the requirements in Eq.(\ref{rl4}), one can achieve
several solutions. They are
\begin{eqnarray}\label{M12}
     &&  |V_{1}|=\frac{1}{\sqrt{3}}\left(\begin{array}{ccc}
       1 & 1 & 1\\
       1 & 1 & 1\\
      1 & 1 &1
      \end{array}\right), |V_{2}|=\left(\begin{array}{ccc}
       {\rm sin}\theta & 0 & {\rm cos}\theta\\
       0 & 1 & 0\\
      {\rm cos}\theta & 0 & {\rm sin}\theta
      \end{array}\right), |V_{3}|=\left(\begin{array}{ccc}
       {\rm sin}\theta & \frac{{\rm cos}\theta}{\sqrt{2}} & \frac{{\rm cos}\theta}{\sqrt{2}}\\
       \frac{{\rm cos}\theta}{\sqrt{2}} & {\rm sin}\theta & \frac{{\rm cos}\theta}{\sqrt{2}}\\
      \frac{{\rm cos}\theta}{\sqrt{2}} & \frac{{\rm cos}\theta}{\sqrt{2}} & {\rm sin}\theta
      \end{array}\right),\nonumber\\&&|V_{4}|=\left(\begin{array}{ccc}
       {\rm sin}\phi& {\rm sin}\phi &  \sqrt{ {\rm cos}2\phi}\\
      \frac{{\rm cos}\phi}{\sqrt{2}} & \frac{{\rm cos}\phi}{\sqrt{2}} & {\rm sin}\phi \\
   \frac{{\rm cos}\phi}{\sqrt{2}} & \frac{{\rm cos}\phi}{\sqrt{2}}& {\rm sin}\phi
      \end{array}\right),|V_{5}|=\left(\begin{array}{ccc}
       {\rm sin}\phi& \frac{{\rm cos}\phi}{\sqrt{2}} & \frac{{\rm cos}\phi}{\sqrt{2}}\\
       {\rm sin}\phi& \frac{{\rm cos}\phi}{\sqrt{2}} & \frac{{\rm cos}\phi}{\sqrt{2}}\\
     \sqrt{ {\rm cos}2\phi}& {\rm sin}\phi & {\rm sin}\phi
      \end{array}\right)
\end{eqnarray}
where $\theta$ lies in the range of $0^\circ\sim 90^\circ$, $\phi$
stays in the range $0^\circ\sim 45^\circ$
and $|V_a|\, (a=1\sim 5)$
represent the mixing matrices which only contain the module of matrix elements. Definitely, in
such a way, the unitarity of the matrix does not manifest at all. Later, see below, when we discuss the practical CKM or PMNS
matrices, we need to input phases by hand. As stated above, as other constraints involving the orthogonality among
the elements are applied, the phases would be automatically taken in, but the procedure for obtaining solutions
 is much more complicated and tedious, so we will leave the task as the goal of our next work. One may notice that $|V_1|$,
$|V_2|$ and $|V_3|$ are just real symmetrical matrixes and $|V_5|$ is
just the transposed matrix of $|V_4|$.


\subsection{Issues related with CKM matrix }
As  $\theta$ in $|V_2$ and $|V_3$ is set to be $90^\circ$, one immediately obtains
\begin{equation}\label{M8p1}|V_{2}|=|V_{3}| =\left(\begin{array}{ccc}
       1 & 0 & 0\\
       0 & 1 & 0\\
      0 & 0 &1
      \end{array}\right)\end{equation}
which is just the the CKM matrix under the limits of
$\theta_{a2}\rightarrow0$ and $\theta_{a3}\rightarrow0$. At this
moment one may be convinced that the way for obtaining solutions
discussed in subsection A is indeed practical. Actually, it is
longtime noticed that the CKM matrix is close to a unit one, and in
Ref.\cite{delAguila:1987se} the authors suggested to transform an
unit matrix to practical CKM by introducing a new $D$ quark.

\subsection{Issues related with some symmetrical PMNS pattern }
Next, let us explore whether these solutions can be related to
the symmetrical PMNS textures.

If $\phi=45^\circ$ in $|V_4|$ we can get
\begin{equation}\label{M8p2}
|V_{4}|=\left(\begin{array}{ccc}
     \frac{ 1 }{\sqrt{2}}& \frac{1}{\sqrt{2}} & 0\\
      \frac{1}{2} & \frac{1}{2} & \frac{1}{\sqrt{2}}\\
    \frac{1}{2} & \frac{1}{2} & \frac{1}{\sqrt{2}}
     \end{array}\right) \end{equation}
which is nothing more, but the modula of the bimaximal mixing
pattern\cite{Vissani:1997pa,Baltz:1998ey,Barger:1998ta}. It is
not astonished because the
proposed PMNS textures satisfy  the equations exactly due to existence of a hidden symmetry.

Cabibbo\cite{Cabibbo:1977nk} and
Wolfenstein\cite{Wolfenstein:1978uw} proposed a symmetrical PMNS
matrix as
\begin{equation}\label{M9}
      V_{CW}=\frac{1}{\sqrt{3}}\left(\begin{array}{ccc}
       1 & 1 & 1\\
       1 & \omega & \omega^2\\
      1 & \omega^2 &\omega
      \end{array}\right),
\end{equation}
where $\omega=e^{i2\pi/3}$. It is found that if only the modules of the matrix elements
are concerned,  the $ |V_{CW}|$ (i.e. as one only keeps the modules of elements) is
just our solution $|V_1|$.  In the $A_4$\cite{Babu:2002dz,Ma:2001dn} or $S_4$
\cite{Hagedorn:2006ug,Ma:2005pd} models,
the charged lepton mass matrix is diagonalized by the unitary $V_{CW}$ and the
Majorana
mass matrix of neutrinos is diagonalized by $V_{\nu}$ which is written as
 \begin{equation}\label{M10}
      V_{\nu}=\left(\begin{array}{ccc}
       \frac{1}{\sqrt{2}} & 0 & - \frac{1}{\sqrt{2}}\\
       0 & 1 & 0\\
      \frac{1}{\sqrt{2}} & 0 & \frac{1}{\sqrt{2}}
      \end{array}\right),
\end{equation}
where $|V_{\nu}|$ is equal to our $|V_2|$ by setting
$\theta=\frac{\pi}{4}$. As we introduce phases in $|V_2|$ to make
it to be $V_2$, then moving further one can obtain the
tribimaximal texture ($V_{TB}$) which is  the product
$V_{CW}V_{\nu}$\cite{Altarelli:2005yx,He:2006dk},
\begin{equation}\label{M3}
      V_{TB}=\left(\begin{array}{ccc}
        \sqrt{\frac{2}{3}} &\frac{1}{\sqrt{3}} &0 \\
         -\frac{1}{\sqrt{6}} &  \frac{1}{\sqrt{3}} &  \frac{1}{\sqrt{2}}\\
          \frac{1}{\sqrt{6}} & -\frac{1}{\sqrt{3}} & \frac{1}{\sqrt{2}}
      \end{array}\right).
  \end{equation}

In Ref.\cite{Qu:2013aca} the authors constructed a new mixing pattern
for neutrinos based on the $\mu-\tau$ interchange symmetry, the
trimaximal mixing in $\nu_2$ and the self-complementarity
relation. The mixing matrix is
\begin{equation}\label{M10}
      V_{QM}= \left(\begin{array}{ccc}
   \frac{ \sqrt{2}+1}{3} & \frac{1}{\sqrt{3}} &- \frac{ \sqrt{2}-1}{3}\\
      -\frac{ \sqrt{2}+1}{6}\mp i\frac{ \sqrt{6}-\sqrt{3}}{6}& \frac{1}{\sqrt{3}} &
      \frac{ \sqrt{2}-1}{6}\mp i\frac{ \sqrt{6}+\sqrt{3}}{6}\\
     \frac{ \sqrt{2}+1}{6}\mp i\frac{ \sqrt{6}-\sqrt{3}}{6} &- \frac{1}{\sqrt{3}} &
     -\frac{ \sqrt{2}-1}{6}\mp i\frac{ \sqrt{6}+\sqrt{3}}{6}
      \end{array}\right).
\end{equation}

In analog to the procedure of obtaining the
tribimaximal mixing pattern  we can derive
$V_{QM}$ from our solutions $V_{1}$ and $V_{2}$, while proper phases are set by hand. Namely, as one sets
 \begin{equation}\label{M10}
      V_{1}=\frac{1}{\sqrt{3}}\left(\begin{array}{ccc}
      1 & 1 & 1\\
       a & 1 & a^*\\
       -a^* & -1 &  -a
      \end{array}\right), V_{2}=\left(\begin{array}{ccc}
       \frac{\sqrt{2}}{\sqrt{3}} & 0 &  \frac{1}{\sqrt{3}}\\
       0 & 1 & 0\\
      \frac{1}{\sqrt{3}} & 0 & -\frac{\sqrt{2}}{\sqrt{3}}
      \end{array}\right),
\end{equation}
the product $V_{1}V_{2}$  just arises the mixing matrix with
$a=-\frac{1}{2}\pm\frac{\sqrt{3}i}{2}$. All the relations between
the solutions with the proposed PMNS were unexpected before.

In Ref.\cite{Yao:2015dwa} the authors assumed that neutrinos
are Dirac particles, then they derived lepton mixing matrices
from the flavor $SU(3)$. We notice that the solutions in
Eq.(\ref{M8p1}), Eq.(\ref{M8p2}) and Eq.(\ref{M9}) can also  be
produced from $SU(3)$ group.

It is noted that  $|V_{1}|$ and $|V_{2}|$ are the solutions of the
equalities but $|V_{TB}|$ and $|V_{QM}|$ are not,
they deviate from the solutions slightly. In Tab.\ref{tab:value} we calculate and list the
deviations of the corresponding quantities in $|V_{TB}|$ and $|V_{QM}|$ from the left sides of Eq.(5). There are five
equalities in Eq.(5), one can notice that a few of the five are satisfied, while the others decline slightly.

\begin{table}
\caption{The values of the left sides in Eq.(5). The labels No.1
to No.5 refer to the first, second, ... equations in the equation group (5).}
\label{tab:value}
\begin{tabular}{cccccc}\hline\hline
  &No.1  &No.2 &No.3&No.4&No.5
 \\\hline
 $|V_{TB}|$&0   &0 &0&0.1 &0.0707 \\
$|V_{QM}|$&0.0020   &-0.0023 &0.0017 &0.0900 &0.0630  \\
\hline\hline
\end{tabular}
\end{table}

\section{SUMMARY AND DISCUSSIONS}

Based on the observed relations ${\rm sin}\delta_a\approx {\rm
sin}\delta_e,\, {\rm sin}\delta_b\approx {\rm sin}\delta_c,\, {\rm
sin}\delta_d\approx {\rm sin}\delta_e,\, {\rm sin}\delta_f\approx
{\rm sin}\delta_h,\, {\rm sin}\delta_h\approx {\rm sin}\delta_i$
among the CP phases in the nine parametrization
schemes which were widely discussed in literature, it is
conjectured that they originate from a high symmetry which later breaks by
some mechanisms. Even though, we so far do not know
what the symmetry is and what breaks it, one can be convinced by those equalities
of their existence. Further assuming those relations to hold exactly, we are able to
establish several scheme-independent
equalities\cite{Ke:2014hxa}. These relations which we obtained by exploring the CKM matrix
also work for the
PMNS matrix. How to understand these relations is one of the task of our work.

This probably corresponds to Lam's suggestion\cite{Lam:2011ip}
that a generic potential is invariant under $U(1)\times SO(3)$ and
the potential causes a breakdown into three phases: the phase I
has an $A_4$ symmetry which is suitable for leptonic mixing
whereas the other two phases have symmetries $SO(2)$ and
$Z_2\times Z_2$. The $SO(2)$ phase is ruled out by phenomenology
and the $Z_2\times Z_2$ is for the quark mixing. We derive similar
results from solving the equalities, i.e. as we showed in
subsections III-C and III-D, $|V_2|$  and $|V_3|$ correspond to
the quark mixing and $|V_4|$ is related to leptonic mixing, and
$|V_2|$ $|V_3|$ and $|V_4|$ all are solutions of Eq.(5). So far,
we have derived the relations and got some symmetric textures from
phenomenology and have not associated the results with the
underlying symmetry yet as discussed above, but we will in our
later works.

It is able to derive similar relations from different starting
points.
When conjectured that these equalities we derived above
can just be the consequences of the small  mixing angles between
quarks, namely irrelevant to any symmetry. Even though these
equalities can be deduced by enforcing certain rephasing
invariants to the quark mixing matrix plus a condition of small
mixing angles, the fact that these equalities also hold for lepton
sector with two mixing angles being sufficiently large, obviously
does not fit the arguments.

We obtain the solutions when the $``\approx"$ sign is set into
$``="$ for those equalities. There are infinite numbers of
solutions and each of them has one free parameter. We note that the unit matrix is also one of the
solutions which is just the limit case of the CKM matrix under the
condition $\theta_2\rightarrow 0$ and $\theta_3\rightarrow 0$ in
any parametrization schemes. It implies that these equalities are
indeed non-trivial after all.

We extend the relations to the lepton case, namely one can
immediately relate some of the obtained solutions to the symmetric
textures for the PMNS
matrices proposed in literatures, such as bimaximal and
tri-bimaximal mixing pattern. Concretely, the bimaximal texture
corresponds to one solution whereas the tri-bimaximal texture can
be related to two solutions.

A more complex mixing texture which was suggested
in Ref.\cite{Qu:2013aca} can be constructed from two of our solutions.
The relations seem to weave a net to include many unexpected phenomena, all these may indicate
that these equalities reflect existence of a
definite symmetry. These equalities may
hold initially at high energy scales, such as the see-saw or GUTs, then the symmetry is distorted or broken
by some mechanisms, and these equalities become
approximate for the CKM and PMNS matrices at practical energy scale.
Further studies on these relations will definitely
lead to eventually understand the symmetry and breaking mechanism.

\section*{Acknowledgement}

We benefit from the illuminating seminar about the underlying symmetry given by Prof. Lam at Nankai University.
This work is supported by the National Natural Science Foundation
of China (NNSFC) under the contract No. 11375128 and 11135009.

\appendix

\section{Check of the relation under limit}
The relations in Eq.(\ref{rl4}) can be proved under some limits.
As an example we check the first one in P$_a$ parametrization.
The left side
\begin{equation}
\lim_{a2\rightarrow 0,a3\rightarrow 0}\frac{{|V_{21}|} {|V_{22}|}
}{
   1-{|V_{23}|}^2}=\lim_{a2\rightarrow 0,a3\rightarrow 0}
   \frac{| -c_{a1}s_{a2}s_{a3} -s_{a1}c_{a2}e^{-i \delta_a}|
   | -s_{a1}s_{a2}s_{a3}+c_{a1}c_{a2}e^{-i\delta_a}|}
   {1-|s_{a2}c_{a3}|^2}=s_{a1}c_{a1}
   \end{equation}

and the right side
\begin{equation}
\lim_{a2\rightarrow 0,a3\rightarrow 0}\frac{{|V_{11}|} {|V_{12}|}
   }{
    {|V_{23}|}^2+{|V_{33}|}^2}=\lim_{a2\rightarrow 0,a3\rightarrow 0}
    \frac{|c_{a1}c_{a3}||  s_{a1}c_{a3}|}{|s_{a2}c_{a3}|^2+|c_{a2}c_{a3}|^2}=s_{a1}c_{a1}
   \end{equation}
so one obtains $\frac{{|V_{21}|} {|V_{22}|} }{
   1-{|V_{23}|}^2}=\frac{{|V_{11}|} {|V_{12}|}
   }{
    {|V_{23}|}^2+{|V_{33}|}^2}$.

\section{Deduction of the relations using the rephasing
invariants of quark mixing matrix}

In P$_a$ parametrization

\begin{equation}
V_{12}  V^*_{22}
V_{23}V^*_{13}=c_{a1}c_{a2}c_{a3}^2s_{a1}s_{a2}s_{a3}e^{i\delta_a}-c_{a3}^2s_{a1}^2s_{a2}^2s_{a3}^2.
\end{equation}

In P$_e$ parametrization

\begin{equation}
V_{12}  V^*_{22}
V_{23}V^*_{13}=-c_{e1}c_{e2}^2c_{e3}s_{e1}s_{e2}s_{e3}e^{-i\delta_e}-c_{e1}^2c_{e2}^2s_{e2}^2s_{e3}^2
\end{equation}

From Eq.(\ref{M1}) and Eq.(\ref{M2}) one has
$s_{a3}=c_{e2}s_{e3}$, $s_{a2}c_{a3}=s_{e2}$,
$c_{a2}c_{a3}=c_{e2}c_{e3}$ so Eq. (B2) changes into
\begin{equation}
V_{12}  V^*_{22}
V_{23}V^*_{13}=-s_{e1}c_{e1}c_{a3}^2s_{a3}s_{a2}c_{a2}e^{-i\delta_e}-c_{e1}^2s_{a3}^2s_{a2}^2c_{a3}^2
\end{equation}

Using the invariants $V_{12}  V^*_{22} V_{23}V^*_{13}$ which is supposed to be free of parametrization schemes,
one can obtain
\begin{equation}
c_{a1}s_{a1}c_{a3}^2s_{a2}s_{a3}c_{a2}e^{i\delta_a}-c_{a3}^2s_{a1}^2s_{a2}^2s_{a3}^2
=-s_{e1}c_{e1}c_{a3}^2s_{a3}s_{a2}c_{a2}e^{-i\delta_e}-c_{e1}^2s_{a3}^2s_{a2}^2c_{a3}^2.
\end{equation}
Dividing it by $c_{a3}^2s_{a2}s_{a3}c_{a2}$
\begin{equation}
c_{a1}s_{a1}e^{i\delta_a}-\frac{s_{a1}^2s_{a2}s_{a3}}{c_{a2}}
=-s_{e1}c_{e1}e^{-i\delta_e}-\frac{c_{e1}^2s_{a3}s_{a2}}{c_{a2}}.
\end{equation}
Considering both the real and imaginary parts to be invariant, and supposing small angles
$\theta_{a2}$ and $\theta_{a3}$, one has  $\tan \delta_a=-\tan
\delta_e$ then the result $\sin \delta_a=\sin \delta_e$ can be
deduced. That is the same as we have by phenomenology which is directly related to experimental
measurements.


\end{document}